\documentstyle[galley, rotate]{mn}
\onecolumn
\input{epsf}
\def\n{{\noindent}}

\def\inc{{\int_0^{\chi_s}}}
\def\ex{{\int {d^2 {\bf l} \over (2
\pi)^2}~ {\rm P} { \big ( {l\over r(\chi)} \big )} W_2^2(l\theta_0)}}
\def\av{\langle \kappa^2(\theta_0) \rangle}
\def\kmin{{\int_0^{\chi_s} \omega(\chi) d \chi }}

\title[Weak Lensing]
{The Statistics of Weak Lensing at Small Angular Scales: \\
Probability Distribution Function}

\author[D. Munshi et al.]{Dipak Munshi$^{1}$
and Bhuvnesh Jain$^2$\\ 
$^1$Max-Planck-Institut fur Astrophysik,
Karl-Schwarzschild-Str.1, D-85740, Garching, Germany\\
$^2$Johns Hopkins University, Department of Physics,
Baltimore, MD 21218, USA}

\pagerange{000,000}
\begin{document}

\maketitle

\begin{abstract}
Weak gravitational lensing surveys have the potential to directly
probe mass density fluctuation in the universe. Recent studies
have shown that it is possible to model the statistics of the convergence
field at small angular scales by modeling the statistics of
the underlying density field in the highly nonlinear regime. We propose a 
new method to  model the complete probability distribution function of the 
convergence field as a function of smoothing angle and source redshift. 
The model relies on a hierarchical ansatz for the 
behavior of higher order correlations of the density field. 
We compare our results with ray tracing simulations and
find very good agreement over a range of smoothing angles. 
Whereas the density probability distribution 
function is not sensitive to the cosmological model, 
the probability distribution function for the convergence
can be used to constrain both the power spectrum and 
cosmological parameters. 
%We also provide an accurate analytical relation
%between cumulants measured from N-Body simulations with that derived from
%ray tracing experiments. 
\end{abstract}


\section{Introduction}

Weak distortions in the images of high red-shift galaxies by the intervening 
large-scale structure provide us with valuable information about the
mass distribution in the universe. The study of such
distortions provides us a unique way to probe the statistical 
properties of the intervening density distribution. 
Traditionally, the study of
gravitational clustering in the quasi-linear and non-linear regime
are done by analyzing galaxy catalogs. However such studies can only
provide us with information on how the galaxies are clustered, and 
to infer the statistics of the underlying mass distribution from 
galaxy catalogs one needs a prescription 
for galaxy bias. Weak lensing studies are free from such effects 
and can directly probe the statistics of the underlying mass distribution.

Pioneering  works in this direction were done by Blandford et
al. (1991), Miralda-Escude (1991) and Kaiser (1992) based on the 
earlier work by Gunn (1967).  Current progress in weak lensing
can broadly be divided into two distinct categories. Whereas Villumsen
(1996), Stebbins (1996), Bernardeau et al. (1997), Jain \& Seljak
(1997) and Kaiser (1998) 
have focussed on the linear and quasi-linear regime by assuming a 
large smoothing angle, several authors have developed a numerical 
technique to simulate weak lensing catalogs. Numerical
simulations of weak lensing typically employ N-body simulations,
 through which ray tracing experiments are conducted
(Schneider \& Weiss 1988; Jarosszn'ski et al. 1990; Lee \& Paczyn'ski
1990; Jarosszn'ski 1991; Babul \& Lee 1991;  Bartelmann \& Schneider
1991,Blandford et al. 1991). Building on the earlier work of Wambsganns
et al. (1995, 1997, 1998) detailed numerical study of lensing 
was done by  Wambsganns, Cen \& Ostriker (1998). Other recent studies
using ray tracing experiments have been conducted by Premadi, Martel
\& Matzner (1998), van Waerbeke, Bernardeau \& Mellier (1998),
Bartelmann et al (1998), Couchman, Barber \& Thomas (1998) and
White \& Hu (1999). 
While  a 
peturbative analysis can provide valuable information at large
smoothing angles, it can not be used to study lensing 
at small angular scales as the whole perturbative series starts to
diverge. 

A complete analysis of weak lensing statistics at small angular scales
is not available at present, as we do not have a similar analysis
for the underlying dark matter distribution. However there are several 
non-linear {\em ansatze} which predicts a tree hierarchy for matter correlation
functions and are thought to be successful to some degree in modeling
results from numerical simulations. Most of these {\em ansatze}
assumes a tree hierarchy for higher order correlation functions and 
they disagree with each other by the way they assign weights to
trees of same order but of different topologies (Balian \& Schaeffer 1989,
Bernardeau \& Schaeffer 1992; Szapudi \& Szalay). The evolution of the
two-point correlation functions in all such approximations 
are left arbitrary. However recent studies by several authors 
(Hamilton et al 1991; Peacock \& Dodds 1994; Nityanada \& Padmanabhan 1994;   
Jain, Mo \& White 1995; Peacock \& Dodds 1996) have provided us 
accurate fitting formulae for the
evolution of the two-point correlation function, which can be used in
combination with these hierarchical {\em ansatze} to predict the
clustering properties of dark matter distribution in the universe.

Most recent studies in weak lensing have mainly focussed on 
lower order cumulants (van Waerbeke, Bernardeau \& Mellier 1998;
Schneider et al 1998; Hui 1999; Munshi, Jain \& White 1999a), 
cumulant correlators (Munshi, Jain \& White 1999a) and errors 
associated with their measurement from observational data using 
different filter functions (Reblinsky et
al. 1999) . However it is well known that higher order 
moments are increasingly 
sensitive to the tail of the distribution function 
and are more sensitive to measurement errors due to the
finite size of a catalog. On the other hand numerical simulations 
involving ray tracing experiments have already shown that while the
probability distribution function associated with the density field is
not sensitive to cosmological parameters, its weak lensing counterpart 
can help us in the estimation of such parameters
from observational data (Jain, Seljak \& White 1999). At present there
is no prescription for the theoretical estimation of the PDF for the smoothed 
convergence field $\kappa(\theta_0)$. Valageas (1999) has
used a hierarchical ansatz for computing the error involved in 
the estimation of $\Omega_0$ and $\Lambda$ from
SNeIa observations. A similar fitting function was recently proposed by
Wang (1999). Our formalism is similar to that of Valageas (1999)  
although the results obtained by us include the effect of 
smoothing. The formalism we introduce can be easily extended
to the multi-point PDF and hence to compute bias and $S_N$ parameters 
associated with ``hot spots'' of convergence maps. Analytical results
and  detailed comparison against numerical simulations will be presented
elsewhere.

The paper is organized as follows. In section $2$ we briefly
the ray tracing simulations. Section $3$ presents the analytical 
results necessary to compute the PDF of the smoothed convergence field
$\kappa(\theta_0)$. The comparison of these results with ray tracing
simulations is made in section $4$. Section $5$ contains a discussion of 
our results.

\section{Generation of Convergence Maps from N-body Simulations}

Convergence maps are generated by solving the geodesic equations for
the propagation of light rays through N-body simulations of dark matter
clustering. The simulations used for our study are adaptive 
$P^3M$ simulations with
$256^3$ particles and were carried out using codes kindly made
available by the VIRGO consortium. These simulations can resolve
structures larger than $30h^{-1}kpc$ at $z = 0$ accurately. These
simulations were carried out using 128 or 256 processors on CRAY T3D
machines at Edinburgh Parallel Computer Center and at the Garching
Computer Center of the Max-Planck Society. These simulations were
intended primarily for studies of the formation and clustering of
galaxies (Kauffmann et al 1999a, 1999b; Diaferio et al 1999) but were
 made available by these authors and by the Virgo
Consortium for this and earlier studies of gravitational lensing 
(Jain, Seljak \& White 1999, Reblinsky et al. 1999,  Munshi \& Jain 1999a)

Ray tracing simulations were carried out by Jain et al. (1999) using
a multiple lens-plane calculation which implements the discrete
version of recursion relations for mapping the photon position and the
Jacobian matrix (Schneider \& Weiss 1988; Schneider, Ehler \& Falco
1992). In a typical experiment $4\times 10^6$ rays are used to trace 
the underlying mass distribution.
The dark matter distribution 
between the source and the observer is projected onto $20$ - $30$ planes.
The particle positions on each plane are interpolated onto a $2048^2$ 
grid. On each plane the shear matrix is computed on this grid 
from the projected density by using Fourier space relations between the
two. The photons are propagated starting from a rectangular grid on the
first lens plane. The regular grid of photon position gets distorted 
along the line of sight. To ensure that all photons reach the
observer, the ray tracing experiments are generally done backward in time
from the observer to the source plane at red-shift $z = z_s$. 
The resolution of the convergence maps is limited by 
both the resolution scale associated with numerical simulations and also 
due to the finite resolution of the grid to propagate photons. 
The outcome of these simulations are shear and 
convergence maps on a two dimensional grid. Depending on the 
background cosmology the two dimensional box 
represents a few degree scale patch on the sky. For more details on the
generation of $\kappa$-maps, see Jain et al (1999).

\section{Analytical Predictions}
In this section we provide the necessary theoretical background for
the analytical derivation of the probability distribution function for the
smoothed convergence field $\kappa(\theta_0)$. We will use models of
nonlinear hierarchical clustering; details of various hierarchical 
ansatzes can be found
in Balian \& Schaeffer (1989), Bernardeau \& Schaeffer (1992,1999),
Bernardeau (1992,1994), Szapudi \& Szalay (1993), Munshi et
al. (1999a,b,c), Valageas \& Schaeffer (1998), Valageas et al. (1997)).
 
The following line element describes the
background geometry of the universe:
\begin{equation}
d\tau^2 = -c^2 dt^2 + a^2(t)( d\chi^2 + r^2(\chi)d^2\Omega)
\end{equation}
Where we have denoted the angular diameter distance by $r(\chi)$ and
scale factor of the universe by $a(t)$; $r(\chi)= K^{-1/2}\sin (K^{-1/2}
\chi)$ for positive curvature, $r(\chi) = (-K)^{-1/2}\sinh
((-K)^{-1/2}\chi)$ for negative curvature and $r(\chi) = \chi$ for a zero
curvature universe. The curvature $K$ is given in terms of the 
present value $H_0$
and $\Omega_0$ as $K= (\Omega_0 -1)H_0^2$. Parameters 
charecterising different cosmological models that we will be studying
are listed in Table 1.

\begin{table}
\begin{center}
\caption{Cosmological parameters  characterizing different models}
\label{tabsig2D}
\begin{tabular}{@{}lcccc}
                  &SCDM&TCDM&LCDM&OCDM\\
$\Gamma$&0.5&0.21&0.21&0.21\\
$\Omega_0$&1.0&1.0&0.3&0.3 \\
$\Lambda_0$&0.0&0.0&0.7&0.0 \\
$\sigma_8$&0.6&0.6&0.9&0.85\\
$H_0$&50&50&70&70\\
\end{tabular}
\end{center}
\end{table}

\subsection{Formalism}

The statistics of the convergence $\kappa$ in weak lensing
is very similar to that of the projected
density. In our analysis we will consider a small patch
of the sky where we can use the plane parallel approximation or small
angle approximation to replace spherical harmonics by Fourier modes.
The 3-dimensional density 
contrast $\delta$ along the line of sight when projected onto the
sky with the weight function $\omega(\chi)$ gives 
the convergence in a direction $\gamma$.

\begin{equation}
\kappa({\bf \gamma}) = \inc {d\chi}
\omega(\chi)\delta(r(\chi){\bf \gamma})
\end{equation}

\n
In our discussion  we will be assume the source galaxies to be at a fixed
 red-shift $z_s$. For this case, one can express the weight function 
as, 
\begin{equation}
\omega(\chi) = {3\over 4} {H_0^2\over c^2}\ 
\Omega_m a \ {r(\chi) r(\chi_s - \chi)\over r(\chi_s)} , 
\end{equation}
where $\chi_s$ is the comoving radial
distance to the source galaxies. Using the Fourier transform of
$\delta$, the convergence can be expressed as:

\begin{equation}
\kappa(\gamma) = \inc {d\chi} \omega(\chi) \int {d^3{\bf k} \over {(2
\pi)}^3} \exp ( i \chi k_{\parallel} + i r \theta k_{\perp} ) \delta_k ,
\end{equation}

\n
where $\theta$ denotes the angle between
the line of sight direction ${\bf \gamma}$ and the wave vector ${\bf
k}$,  $k_{\parallel}$ and $k_{\perp}$ denote the components
of ${\bf k}$, parallel and perpendicular to the line of sight.
In the small angle approximation one assumes that $k_{\perp}
>> k_{\parallel}$. 
Using these definitions we can compute the
projected two-point correlation 
function in terms of the dark matter power spectrum 
$P(k,\chi)$ (Peebles 1980, Kaiser 1992, Kaiser 1998):

\begin{equation}
\langle \kappa(\gamma_1) \kappa(\gamma_2) \rangle = \inc d {\chi}
{\omega^2(\chi) \over r^2(\chi)} \int {d^2 {\bf l} \over (2
\pi)^2}~\exp ( \theta l )~ {\rm P} { \big ( {l\over r(\chi)}, \chi \big )}.
\end{equation}

\n
Where we have introduced ${\bf l} = r(\chi){\bf
k}_{\perp}$ to denote the scaled wave vector projected on the
sky. The variance of $\kappa$
smoothed over an angle $\theta_0$ is (Jain \& Seljak 1997) 

\begin{equation}
\langle \kappa^2({\theta_0}) \rangle = \inc d {\chi_1}
{\omega^2(\chi_1) \over r^2(\chi_1)} \int {d^2 {\bf l} \over (2
\pi)^2}~ {\rm P} { \Big ( {l\over r(\chi)}, \chi \Big )} W_2^2(l\theta_0)
\label{kappa_variance}
\end{equation}

\n
Similarly the higher order moments of the smoothed convergence
field relate $\langle \kappa^p ({\theta_0} \rangle$ to the
3-dimensional multi-spectra of the 
underlying dark matter distribution $B_p$ (Hui 1999, Munshi \& Coles 1999a):

\begin{equation}
\langle \kappa^3 ({\theta_0}) \rangle = \inc d {\chi_1}
{\omega^3(\chi) \over r^6(\chi)} \int {d^2 {\bf l_1} \over (2\pi)^3}
W_2(l_1 \theta_0) \int {d^2{\bf
l_2}\over (2\pi)^2} W_2(l_2 \theta_0) \int {d^2 {\bf l_3} \over
(2\pi)^3} W_2(l_3 \theta_0) ~ {\rm B}_3 \Big ( {l_1\over r(\chi)},
{l_2\over r(\chi)},  {l_3\over r(\chi)}, \chi \Big )_{\sum {\bf l}_i = 0}
\end{equation}

\begin{eqnarray}
\langle \kappa^4 ({\theta_0}) \rangle = \inc d {\chi_1}
{\omega^4(\chi) \over r^8(\chi)} \int {d^2 {\bf l_1} \over (2\pi)^3}
W_2(l_1 \theta_0) \int {d^2{\bf
l_2}\over (2\pi)^2} W_2(l_2 \theta_0) \int {d^2 {\bf l_3} \over
(2\pi)^2} W_2(l_3 \theta_0)\int {d^2 {\bf l_4} \over 2 \pi^2} W_2(l_4
\theta_0)\\ \nonumber  ~ {\rm B}_4 \Big ( {l_1\over r(\chi)},
{l_2\over r(\chi)}, {l_3\over r(\chi)}, {l_4\over r(\chi)}, \chi \Big )_{\sum {\bf l}_i = 0}.
\end{eqnarray}

\n
We will use these results to show that it is possible
to compute the complete probability distribution function
of $\kappa$ from the underlying dark matter probability
distribution function. Details of the analytical results presented
here can be found in Munshi \& Coles (1999b).

\subsection{Hierarchical {\em Ansatze}}

The spatial length scales corresponding to
small angles are in the highly non-linear regime. Assuming a tree model 
for the matter correlation
hierarchy in the highly non-linear regime, one can write the 
general form of the $N$th order correlation function as 
(Groth \& Peebles 1977, Fry \& Peebles 1978,  Davis \&
Peebles 1983, Bernardeau \& Schaeffer 1992, Szapudi \& Szalay 1993):

\begin{equation}
\xi_N( {\bf r_1}, \dots {\bf r_N} ) = \sum_{\alpha, \rm N-trees}
Q_{N,\alpha} \sum_{\rm labellings} \prod_{\rm edges}^{(N-1)}
\xi({\bf r_i}, {\bf r_j}) .
\end{equation}

It is interesting to note that a similar hierarchy 
develops in the quasi-linear regime in the limit of vanishing variance
(Bernardeau 1992); however the hierarchal amplitudes $Q_{N, \alpha}$
become shape dependent functions in the quasilinear regime. In the 
highly nonlinear 
regime there are some indications that these functions become
independent of shape, as suggested by studies of the
lowest order parameter $Q_3 = Q$ using high resolution numerical
simulations (Sccocimarro et al. 1998). In Fourier space such an
ansatz means that the hierarchy of multi-spectra 
can be written as sums of products of the matter power-spectrum.

\begin{eqnarray}
&&B_2({\bf k}_1, {\bf k}_2, {\bf k}_3)_{\sum k_i = 0} = Q ( P({\bf
k_1})P({\bf k_2}) + P({\bf k_2})P({\bf k_3})
+ P({\bf k_3})P({\bf k_1}) ) \\ \nonumber
&&B_3({\bf k}_1, {\bf k}_2, {\bf k}_3, {\bf k}_4)_{\sum k_i = 0} = R_a
\ P({\bf k_1})P({\bf k_1 +
k_2}) P({\bf k_1 + k_2 + k_3})  + {\rm cyc. perm.} + R_b \ P({\bf
k_1})P({\bf k_2})P({\bf k_3}) + 
{\rm cyc. perm.} \\ \nonumber
\end{eqnarray}

\n
Different hierarchical models differ in the way they predict the
   amplitudes of different tree topologies. Bernardeau \&
Schaeffer (1992) considered the case where amplitudes in general are
factorizable, at each order one has a new ``star'' amplitude 
 and higher order ``snake'' and ``hybrid'' amplitudes can
be constructed from lower order ``star'' amplitudes (see Munshi,
Melott \& Coles 1999a,b,c for a detailed description). In models proposed by
Szapudi \& Szalay (1993) it was assumed that all hierarchal amplitudes of any
given order are degenerate. Galaxy surveys
have been used to study these {\em ansatze}. Our goal here is to
show that weak-lensing surveys can also provide valuable information
in this direction, in addition to constraining the matter power-spectra and
background geometry of the universe. We will use the model proposed by 
Bernardeau \& Schaeffer (1992) and its generalization to the 
quasi-linear regime by Bernardeau (1992, 1994) to construct the PDF
of the convergence field $\kappa(\theta_0)$. We express 
the one-point cumulants as:

\begin{eqnarray}
&&\langle \kappa^3(\gamma) \rangle = (3Q_3){\cal C}_3[\kappa^2_{\theta_0}] 
= S_3 \langle \kappa^2(\theta_0) \rangle^2 \label {hui} \\  
&&\langle \kappa^4(\gamma) \rangle = (12R_a + 4
R_b){\cal C}_4[\kappa^3_{\theta_0}] = S_4 \langle \kappa^2(\theta_0) \rangle^3,
\label{Sn}
\end{eqnarray}

\n
where

\begin{equation}
{\cal C}_t[\kappa^m(\theta_0)] = \int_0^{\chi_s} { \omega^t(\chi)
\over r^2(t-1)(\chi)}\kappa^m_{\theta_0} d\chi,
\end{equation}

\n
and

\begin{equation}
\kappa_{\theta_0} =  \int  \frac{d^2\bf l}{(2\pi)^2} P
\left( {l \over r(\chi)} \right) W_2^2(l \theta_0).
\end{equation}

\noindent
Equation ({\ref {hui}}) was derived by Hui (1998). He 
showed that his result agrees well with the ray tracing
simulations of Jain, Seljak and White (1998). More recent studies
have shown that two-point statistics such as cumulant correlators 
can also be modeled in a similar way (Munshi \& Coles 1999, Munshi, 
Jain \& White 1999).

%\begin{figure}
%\protect\centerline{
% \epsfysize = 3.8truein
% \epsfbox[57 126 477 564]
% {outs.ps} }
% \caption{Convergence or $\kappa$ map generated from N-body Simulation
%for SCDM model}
%\end{figure}

\subsection{Scaling Models and Generating Functions}

The success of analytical results on lower order cumulants, when compared
with numerical ray tracing 
experiments, motivates a more general analysis of the probability
distribution function of the smoothed convergence field
$\kappa(\theta_0)$. For this
purpose we found the formalism developed by Balian \&
Schaeffer (1989) and Bernardeau \& Schaeffer (1992) to be most suitable.
These results are based on a general tree hierarchy of higher order
correlation functions and the assumption that the amplitudes associated 
with different tree-topologies are constant in the highly
non-linear regime. These results were generalized by 
Bernardeau (1992, 1994) to the quasi-linear regime 
where the perturbative dynamics can be used to make more concrete
predictions. In this section we review the basic results from the
scaling models before extending such models to the statistics of the
smoothed convergence field $\kappa(\theta_0)$ (for more details see
Munshi \& Coles 1999b). We will use the small
angle approximation throughout our derivation. Our results
can be generalized to the case of the projected clustering of
galaxies. 

In a scaling analysis of the probability distribution function (PDF), the
void probability distribution function (VPF) plays a fundamental role. 
The VPF can be related to the generating function of the cumulants 
or $S_N$ parameters, $\phi(y)$ (White 1979, Balian \& Schaeffer 1989). 
\begin{equation}
P_v(0) = \exp ( \bar N \sigma(N_c) ) = \exp \Big ( - { \phi (N_c) \over
\bar \xi_2} \Big ) ,
\end{equation}
\n
where $P_v(0)$ is the probability of having no ``particles'' in a cell of 
of volume $v$, $\bar N$ is the average occupancy of the particles and 
$N_c = \bar N {\bar \xi_2}$. The VPF is meaningful only for a discrete 
distribution of particles and can not be defined for smoothed 
fields such as $\delta$ or $\kappa(\theta_0)$. However the scaling
functions defined above as
$\sigma(y) = -\phi(y)/y$, are very useful even for
continuous distributions where they can be used as a generating
function of the one-point cumulants or the $S_N$ parameters. 
\begin{equation}
\phi(y) = \sum_{p=1}^{\infty} { S_p \over p! } y^p
\end{equation}
The function $\phi(y)$ satisfies the constraint $S_1 = S_2 = 1$ necessary for
the normalization of the PDF. The other generating function
which plays a very important role in such an analysis is the generating 
function for vertex amplitudes $\nu_n$, associated with nodes appearing in
the tree representation of the higher order correlation hierarchy ($Q_3 =
\nu_2$, $R_a = \nu_2^2$ and $R_b = \nu_3$). 

\begin{equation}
{\cal G}(\tau) = 1 - \tau + { \nu_2 \over 2 ! } \tau^2 - { \nu_3 \over
3! } \tau^3 + \dots
\end{equation}

\n 
A more specific model for ${\cal G}(\tau)$ is sometimes used, which
is useful in making more detailed predictions (Bernardeau \& Schaeffer
1979):

\begin{equation} 
{\cal G}(\tau) = \Big ( 1 + {\tau \over \kappa_a} \Big )^{-\kappa_a}
\end{equation}

\n
We will relate $\kappa$ with other parameters of scaling models.
While the definition of VPF does not use any specific form of the
hierarchical ansatz, writing the tree
amplitudes in terms of the weights associated with the nodes is only
possible when one assumes a factorizable model for the tree hierarchy
(Bernardeau \& Schaeffer 1992). Moreover, in such an ansatz the generating
functions for the tree nodes can be related to the VPF by solving a
pair of implicit equations (Balian \& Schaeffer 1989).

\begin{eqnarray}
&&\phi(y) = y {\cal G}(\tau) - { 1 \over 2} y {\tau} { d \over d
\tau}{\cal G}(\tau) \\
&&\tau = -y { d \over d\tau} {\cal G}(\tau)
\end{eqnarray}

\n
The VPF and PDF can be related to each other by the following equation
(Balian \& Schaeffer 1989)
\begin{equation}
P(\delta) = \int_{-i\infty}^{i\infty} { dy \over 2 \pi i} \exp \Big [ {(
1 + \delta )y - \phi(y)  \over \bar \xi_2} \Big ] \label{pdf}
\end{equation}

\n
So it is clear that the properties of $\phi(y)$ completely determine 
the behavior of $P(\delta)$ for all values of $\delta$. However
different asymptotic expressions of $\phi(y)$ govern the behaviour
of $P(\delta)$ for different intervals of $\delta$. For large $y$ we 
can express $\phi(y)$ as
\begin{equation}
\phi(y) = a y^{ 1 - \omega}.
\end{equation}
We have introduced a new variable $\omega$ for the description of
VPF. This parameter plays a very important role in the scaling analysis.
No theoretical analysis has been done so far to link $\omega$ with
the initial power spectral index $n$. Numerical simulations are generally
used to fix $\omega$ for different initial conditions. Such studies have 
confirmed that an increase of power on smaller scales increases the value
of $\omega$. Typically an initial power spectrum with spectral index $n=
-2$ (which should model CDM-like spectra that we have considered in our
simulations at small length scales) produces a value
of $\omega = 0.3$ which we will use in our analysis of the PDF of 
the convergence field $\kappa(\theta_0)$. 
$\phi(y)$ exhibits a singularity for small but negative value of $y_s$
\begin{equation}
\phi(y) = \phi_s - a_s \Gamma(\omega_s) ( y - y_s)^{-\omega_s}.
\end{equation}

The parameter $k_a$ which we have introduced in the definition of
$\cal G(\tau)$ can be related to the parameters $a$ and $\omega$ appearing 
in the asymptotic expressions of $\phi(y)$ (Balian \& Schaeffer 1989. 
Bernardeau \& Schaeffer 1992).

\begin{figure}
\protect\centerline{
 \epsfysize = 4.truein
 \epsfbox[19 146 589 714]
 {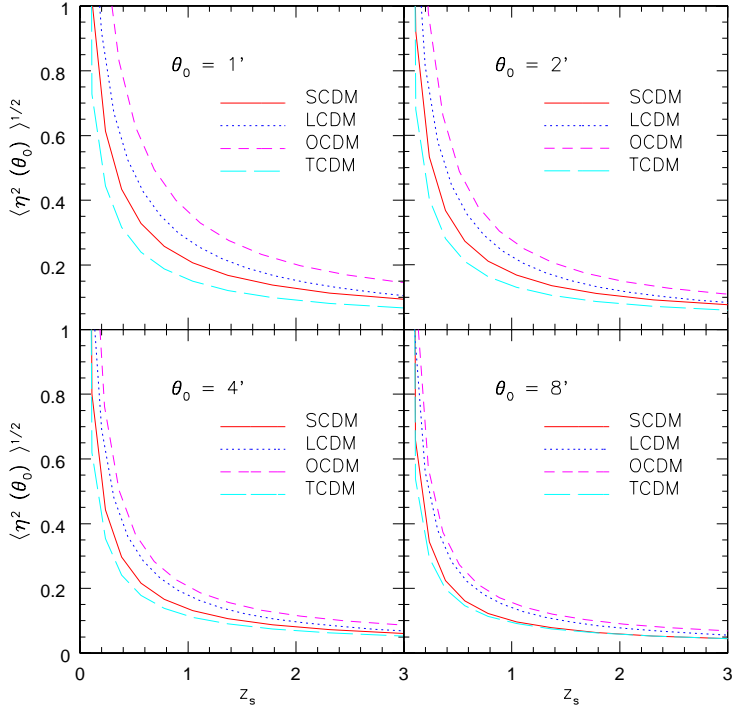} }
 \caption{Analytical predictions for the variance of the reduced
 convergence field $\eta$ as a function of red-shift. Note that
whereas the variance of $\kappa(\theta_0)$ increases with increasing
source red-shift, the trend is reversed for
 $\eta(\theta_0)$. Both $\eta(\theta_0)$ and $\kappa(\theta_0)$ 
decrease with increasing smoothing radius $\theta_0$. }
\end{figure}

\begin{eqnarray}
&&\omega = k_a / ( k_a + 2) \\
&&a = {k_a + 2 \over 2} k_a^{ k_a /  k_a + 2}
\end{eqnarray}

\n
Similarly the parameters $y_s$ and $a_s$ which describe the behavior
of $\phi(y)$ near the singularity can be related to the behavior of
${\cal G(\tau)}$ near its singularity $\tau_s$, which is given by
(Balian \& Schaeffer 1989, Bernardeau \& Schaeffer 1992)

\begin{equation}
\tau_s = {{\cal G}'( \tau_s) \over {\cal G}'(\tau_s) } .
\end{equation}

\n
The parameter $y_s$ can be expressed 
as (Balian \& Schaeffer 1989, Bernardeau \& Schaeffer 1992)

\begin{equation}
y_s = - { \tau_s \over {\cal G}'(\tau_s)} , \\
\end{equation}

\n
which in turn can be related to $k_a$ as

\begin{equation}
-{ 1 \over y_s} = x_{\star} = {1 \over k_a } { (k_a + 2)^{k_a + 2} \over (k_a + 1)^{k_a+1}}.
\end{equation}

\n
These singularities describe $P(\delta)$ for large values of $\delta$.
The newly introduced variable $x_\star$ will be useful later to define 
the large $\delta$ tail of $P(\delta)$. The different asymptotes  in $\phi(y)$
are linked to the behavior of $P(\delta)$ for various values of
$\delta$. However for very large values of the variance i.e. $\bar \xi_2$ 
it is possible to define a scaling function $P(\delta) = h(x)/{\bar \xi}_2^2$  
which will encode 
the scaling behavior of the PDF, where $x$ plays the role of the scaling 
variable and is defined as $(1 + \delta)/\xi_2$. We list the
different ranges of $\delta$ and specify the behavior of $P(\delta)$
in these regimes (Balian \& Schaeffer 1989).

\begin{equation}
{\bar \xi_2 }^{ - \omega \over ( 1 - \omega)} << 1 + \delta << \bar \xi_2;
~~~~~~
P(\delta) = { a \over \bar \xi_2^2} { 1- \omega \over \Gamma(\omega)}
\Big ( { 1 + \delta \over \bar \xi_2 } \Big )^{\omega - 2}
\end{equation}

\begin{equation}
1+ \delta >> {\bar \xi}_2; ~~~~
P(\delta) = { a_s \over \bar \xi_2^2 } \Big ( { 1 + \delta \over \bar
\xi_2}  \Big ) \exp \Big ( - { 1 + \delta \over x_{\star} \bar \xi_2}
\Big )
\end{equation}

\begin{figure}
\protect\centerline{
 \epsfysize = 4.truein
 \epsfbox[19 146 589 714]
 {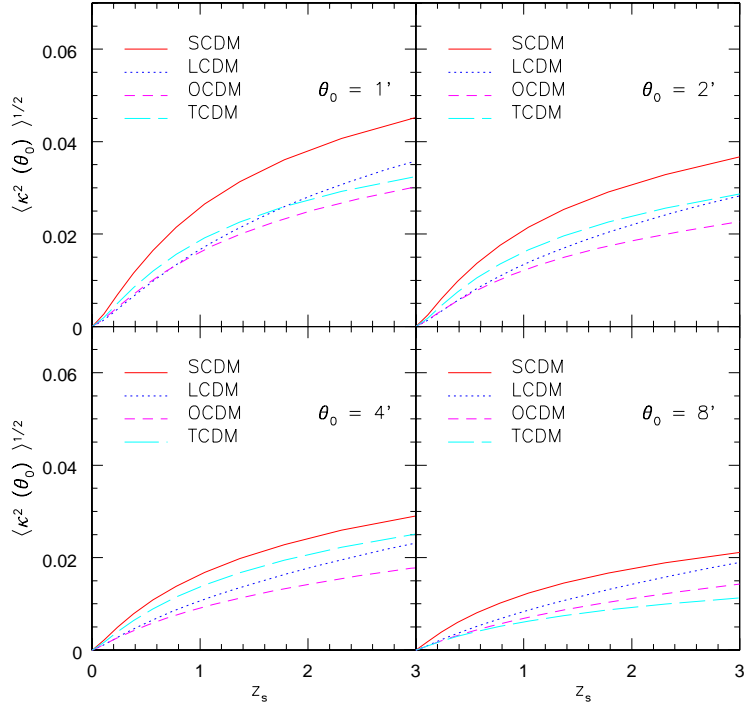} }
 \caption{The variance of the 
convergence field $\langle \kappa^2(\theta_0) \rangle$ as 
a function of red-shift $z_s$. Different panels correspond to 
different smoothing angles.
Note that the ordering of the variance for the convergence field of different 
cosmological  models is not the same as that of the reduced  convergence 
field $\eta$.}
\end{figure}

\n
The integral constraints satisfied by the scaling function are
$ S_1 = \int_0^{\infty} x h(x) dx = 1$ and 
$ S_2 = \int_0^{\infty} x^2 h(x) dx = 1$. These constraints lead to 
the correct normalization of $P(\delta)$. Several numerical
studies have been conducted to study the behavior of $h(x)$ for 
different initial conditions (e.g. Colombi et al. 1996; Munshi et
al. 99). For very small values of $\delta$ the behavior of
$P(\delta)$ is determined by the asymptotic behavior of $\phi(y)$ 
for large values of $y$. Hence 
the PDF can be expressed as (Balian \& Schaeffer 1989):

\begin{equation}
P(\delta) = \int_{-i\infty}^{\infty} { dy \over 2 \pi i} \exp \Big [ {(
1 + \delta )y - a y^{ -\omega} \over \bar \xi_2} \Big ] .
\end{equation}

\n
The above equation implies that it is possible to define another 
scaling function 
$g(z)$, completely determined by
$\omega$, with the scaling parameter $z = (1+
\delta)a^{-1/(1-\omega)}{\bar \xi}_2^{\omega /(1 - \omega)}$. 
However, numerically it is much easier to determine $\omega$ 
from the study of $\sigma(y)$ compared to that from $g(z)$ (e.g. Bouchet
\& Hernquist 1992).

\begin{equation}
1 + \delta << \bar \xi_2;~~~~
P(\delta) = a^{ -1\over (1 - \omega) } {{ \bar \xi}_2 ^{ \omega \over
( 1 - \omega ) }} g(z)
\end{equation}

\begin{equation}
g(z) = \int_{i \infty}^{i \infty} { dt \over 2 \pi i} \exp ( z t - t
^{1- \omega} ) 
\end{equation}

\begin{equation}
P(\delta) = a^{ -1 \over 1 - \omega} {\bar \xi}_2^{ \omega \over 1 -
\omega } \sqrt { ( 1 - \omega )^{ 1/\omega } \over 2 \pi \omega z^{(1
+ \omega)/ \omega } } \exp \Big [ - \omega \Big ( {z \over 1 - \omega}
\Big )^{- {{1 - \omega} \over \omega}} \Big ]
\end{equation}

\n
To summarize, the entire behavior of $P(\delta)$ is
encoded in the two scaling functions, $h(x)$ and $g(x)$. These
functions are relevant for the behavior of $P(\delta)$ at small
and large $\delta$, respectively. Typically $P(\delta)$ shows a cutoff at
both large and small values of $\delta$ and exhibits a
power-law in the middle. The power law behavior is observed when both
$g(z)$ and $h(x)$ overlap, in the highly non-linear
regime and with the decrease in $\bar \xi_2$ the range of $\delta$ 
for which $P(\delta)$ shows such a power law behavior decreases,
finally to vanish in case of small variance i.e. in the
quasi-linear regime.

In the quasi-linear regime exactly the same formalism can be used to
study the PDF. However the generating function now
can be explicitly evaluated by using tree-level perturbative
dynamics (Bernardeau 1992). It is also possible to take smoothing 
corrections into account in which case one can have an explicit
relation of $\omega$ and the initial power spectral index $n$.
In general the parameters $k_a$ or $\omega$ charecterising the VPF
or the CPDF are different from their highly non-linear counterparts.

The PDF now can be expressed in terms of $G_{\delta}(\tau)$
(Bernardeau 1992, Bernardeau 1994):

\begin{eqnarray}
&&P(\delta)\ d \delta = { 1 \over -G_{\delta}'(\tau) } \Big [ { 1 - \tau G_{\delta}''(\tau)
/G_{\delta}'(\tau) \over 2 \pi {\bar \xi}_{2} }  \Big ]^{1/2} \exp \Big ( -{ \tau^2
\over 2 {\bar \xi}_{2}} \Big ) d \tau  \\
&&G_{\delta}(\tau) = {\cal G}(\tau) -1 = \delta 
\end{eqnarray}

\n
The above expression for $P(\delta)$ is valid for $\delta < \delta_c$ 
where $\delta_c$
is the value of $\delta$ which cancels the numerator of the prefactor 
of the exponential function appearing in the above expression. For
$\delta > \delta_c$ the PDF develops an exponential tail which is 
related to the presence of a singularity in $\phi(y)$,
as was the case of its highly non-linear counterpart (Bernardeau
1992, Bernardeau 1994),

\begin{figure}
\protect\centerline{
 \epsfysize = 3.5truein
 \epsfbox[19 431 306 714]
 {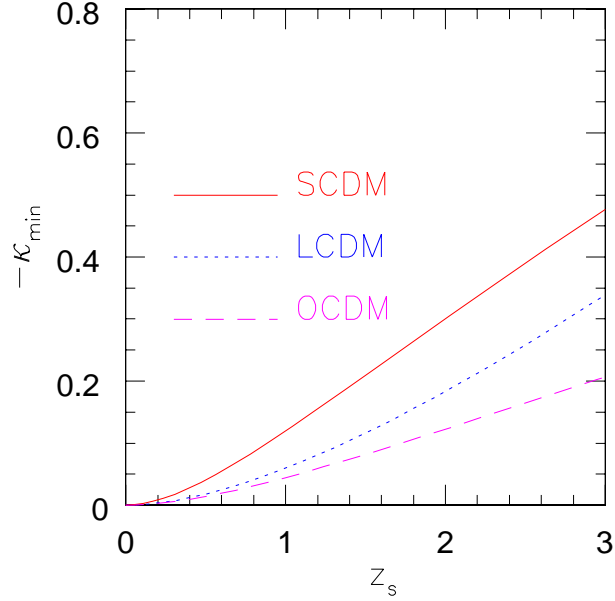} }
 \caption{The minimum value of $\kappa(\theta_0)$, i.e. $-\kappa_{min}$ 
is plotted as a function of the source red-shift $z_s$. 
The minimum value of $\kappa$ occurs in those
line of sight direction where smoothed density perturbation $\delta =
 -1$ all along the line of sight. 
It is independent of the smoothing angle $\theta_0$ and 
depends only on the background geometry. 
}
\end{figure}

\begin{equation}
P(\delta) \ d \delta = { 3 a_s \sqrt {{\bar \xi}_2} \over 4  {\sqrt \pi} }
\delta^{-5/2} \exp \Big [ -|y_s|{ \delta \over {\bar \xi}_{2}} + {|\phi_s|
\over {\bar \xi}_{2}} \Big ] d \delta
\end{equation}

Since the value of the parameter $\omega$ is different in the two 
regimes, the
associated parameters such as $y_s$ and $a_s$ will also be different
in  these two regimes. We will show that the expressions for the
probability distribution function $P(\eta)$ of the reduced convergence
$\eta$ will have exactly the same asymptotes as the weakly
non-linear PDF of $P(1+\delta)$, although the parameters associated
with $P(\eta)$ will correspond to the ones used in the highly non-linear
regime for $P(1+\delta)$. This is one of the important result of our analysis.

It may be noted that similar
expressions can be derived for the approximate dynamics 
sometimes used to simulate gravitational clustering in the weakly
non-linear regime, e.g. Lagrangian perturbation theory which is an extension
of the Zeldovich approximation (Munshi et al. 1994).

\subsection{PDF of the Smoothed Convergence Field} 

For computing the probability distribution function of the smoothed 
convergence field $\kappa(\theta_0)$, we will begin by constructing
its associated cumulant generating function $\Phi_{1+\kappa(\theta_0)}(y)$:

\begin{equation}
\Phi_{1 + {\kappa(\theta_0)}}(y) = y + \sum_{p=2}^ {\infty} {{\langle
\kappa^p(\theta_0) \rangle} \over \langle \kappa (\theta_0 )
\rangle^{p-1}} y^p
\end{equation}

\n
Now using the expressions for the higher moments of the convergence
in terms of the matter power spectrum (equations \ref{kappa_variance} 
and \ref{Sn}) gives,

\begin{equation}
\Phi_{1 + {\kappa({\theta_0})}}(y) = y + \int_0^{\chi_s} \sum_{N=2}^{\infty}
{ (-1)^N \over N!} S_N { \omega^N(\chi) \over r^{2(N-1)} ( \chi)}
\Big [ \ex \Big ]^{ (N-1)} { y^N \over \av^{(N-1)} }. 
\end{equation}

\n
We can now use the definition of $\phi(y)$ for the matter cumulants to
express $\Phi_{1 + \kappa(\theta_0)}(y)$, in terms of $\phi(y)$:

\begin{equation}
\Phi_{1+\kappa_{\theta_0}}(y) =  \int_0^{\chi_s} d \chi
\Big[ {\langle r^2(\chi) \kappa^2_{\theta}  \rangle  \over  \ex} \Big
] \phi \Big [ {\omega (\chi) \over r^2 (\chi)} {\ex \over \av} \Big ]
- y\int_0 ^{\chi_s} \omega(\chi) d \chi
\end{equation}

The extra term comes from the $N=1$ term in the expansion 
of $\Phi_{1+\kappa(\theta_0)}$.
Note that we have used the fully non-linear generating function $\phi$
for the cumulants, though we will use it to construct a generating 
function in the quasi-linear regime.

\n
The analysis becomes much easier if we define a new reduced convergence 
field $\eta(\theta_0)$:

\begin{equation}
\eta({\theta_0}) = {{ \kappa({\theta_0}) - \kappa_{min} } \over
-\kappa_{min}} = 1 + {\kappa({\theta_0}) \over |\kappa_{min}| } ,
\end{equation}

\n
where the minimum value of $\kappa(\theta_0)$ i.e. $\kappa_{min}$ occurs
when the line of sight goes through regions that are completely empty of
matter (i.e. $\delta = -1$ all along the line of sight):

\begin{equation}
\kappa_{min} = - \int_0^{\chi_s} d \chi \omega(\chi) .
\end{equation}

Figures 1 and 2 compare
the variance of $\eta$ and $\kappa$. 
While $\kappa({\theta_0})$ depends on the smoothing
angle, its minimum value $\kappa_{min}$
depends only on the source red-shift and background geometry of the
universe, independent of the smoothing radius. Figure 3 shows the
dependence of $\kappa_{min}$ on source red-shift for three
cosmological models. With the reduced
convergence $\eta$, the cumulant generating function is given by,

\begin{equation}
\Phi_{\eta} (y) = { 1 \over [\kmin]} \int_0^{\chi_s} d \chi 
\Big [{ r^2(\chi) \over \kmin}{ \av \over \ex }\Big ] \phi \Big [
\omega(\chi) \kmin { \ex \over r^2(\chi) \av}   \Big ] 
\end{equation}

%%%\begin{figure}
%%%\protect\centerline{
%%% \epsfysize = 3.5truein
%%% \epsfbox[19 431 306 714]
%%% {k_min.ps} }
%%% \caption{The minimum value of $\kappa(\theta_0)$, i.e. $-\kappa_{min}$ 
%%%is plotted as a function of the source red-shift $z_s$. 
%%%The minimum value of $\kappa$ occurs in those
%%%line of sight direction where smoothed density perturbation $\delta =
%%% -1$ all along the line of sight. 
%%%It is independent of the smoothing angle $\theta_0$ and 
%%%depends only on the background geometry. 
%Note that $-\kappa_{min}$ increases 
%with increase in source red-shift $z_s$ and it takes it maximum value 
%for SCDM  and minimum value for OCDM model for all $z_s$. 
%This is related to the fact
%that the peak height for $P(\kappa(\theta_0))$ is lowest for SCDM and 
%highest for OCDM. Since $-\kappa_{min}$ is same for SCDM and TCDM
% models the difference in analytical as well as simulated
% $P(\kappa(\theta_0)$ is only due to the difference in their initial power
% spectrum. }
%%%}
%%%\end{figure}

\begin{figure}
\protect\centerline{
\epsfysize = 3.truein
\epsfbox[29 400 589 714]
 {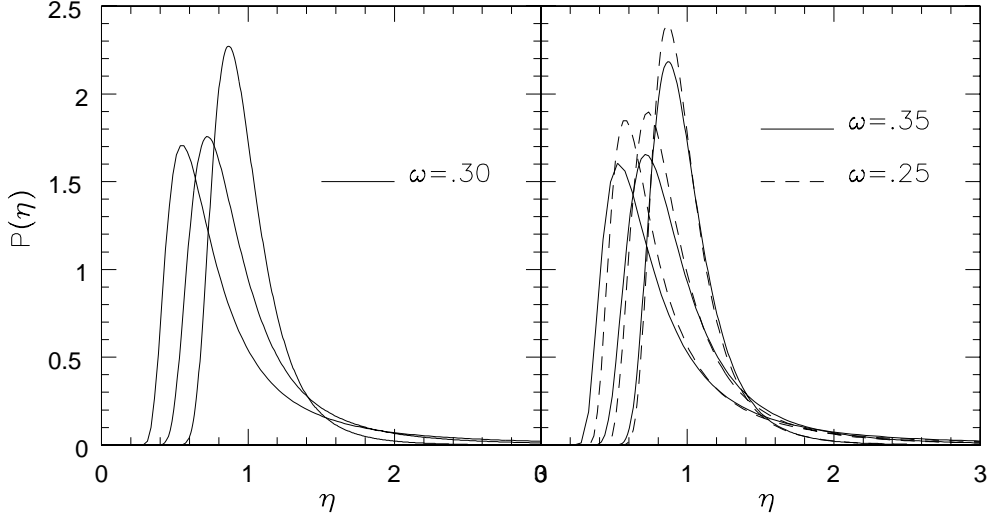} }
 \caption{Analytical predictions for the probability distribution function
$P(\eta)$ as a function of the reduced convergence $\eta$. Curves 
 from left to right 
correspond to decreasing values of the variance ${\sqrt \xi_{\eta}} =
 .25, .5,$ and $ 1.$. In all our comparison with 
ray tracing experiments we have used $\omega = 0.3$. However there 
are some uncertainities in estimation of $\omega$ from numerical
 simulations (see Colombi et. al. 1996; Munshi et al. 1999). In the
 right panel we show how small changes in $\omega$ change the
probability distribution function for the reduced convergence $\eta$.
%$\omega$ depends on initial power spectral index $n$, at present there 
%is no analytical model relating $\omega$ with $n$ and most estimations
%of $\omega$ rely on numerical simulations. Estimations of $\omega$ are
%generally done by studying scaling properties of PDF and VPF and are
% more difficult for initial power spectra with more power on larger
% scales due to the presence of finite volume correction. Notice that
%such finite volume corrections are also present in determination of
% $\kappa$ PDF from ray tracing experiments and generally the affect
%the high $\kappa$ tail of the distribution. }
}
\end{figure}

The constructed cumulant generating function
 $\Phi_{\eta}(y)$ satisfies the normalization constraints $S_1 = S_2 = 1$.
The scaling function associated with $P(\eta)$ can now be 
 easily related with the matter scaling function $h(x)$ 
introduced earlier:

\begin{equation}
h_{\eta} (x) = - \int_{-\infty}^{\infty} { dy \over 2 \pi i} \exp (x
y) \Phi_{\eta} (y) .
\end{equation}

\n
Using this definition we can write:

\begin{eqnarray}
h_{\eta} (x) = { 1 \over [\int_0^{\xi_s} d \chi \omega(\chi)] }
\int_0^{\chi_s} \omega (\chi) d \chi \Big [ {\langle
\kappa_{\theta_0}^2\rangle \over \omega (\chi) \int d^2 {\bf k}_{\perp} P({\bf
k}_{\perp}) W^2( {\bf k}_{\perp} \theta_0 r(\chi)) \int_0^{\chi_s} d \chi \omega
(\chi)} \Big ]^2 \times \\ \nonumber
   h \Big ({\langle
\kappa_{\theta_0}^2\rangle x \over \omega (\chi) \int d^2 {\bf k}_{\perp} P({\bf
k}) W^2( {\bf k}_{\perp} \theta_0 r(\chi)) \int_0^{\chi_s} d \chi \omega
(\chi)}\Big ).  
\end{eqnarray}

\n
While the expressions derived above are exact and are derived for the most
general case using only the small angle approximation, they can be
simplified considerably using further approximations. In the following we will
assume that the contribution to the $\chi$ integrals can be
replaced by an average value coming from the maximum of $\omega(\chi)$,
i.e. $\chi_c$ ($0<\chi_c<\chi_s$). So we replace $\int f(\chi) d\chi$
by $1/2 f(\chi_c)\Delta_{\chi}$ where $\Delta_{\chi}$ is the
interval of integration, and $f(\chi)$ is the function of $\chi$ under
consideration.  Similarly we replace the $\omega(\chi)$ dependence in 
the ${\bf k}$ integrals by $\omega(\chi_c)$. 
Under these approximations we can write,  
 
\begin{eqnarray}
\Phi_{\eta}(y) = \phi(y) \\
h_{\eta}(x) = h(x) 
\end{eqnarray}

\n
Thus we find that the statistics of the underlying field $1+\delta$ and the 
statistics of 
the reduced convergence $\eta$ are exactly the same under such an
approximation (the approximate functions $\Phi_{\eta}$ and
$h_{\eta}(x)$ do satisfy the proper normalization constraints). Although it is
possible to integrate the exact expressions of the scaling functions, 
there is some uncertainty involved in
the actual determination of these functions and 
associated parameters 
such as $\omega, k_a, x_{\star}$ from N-body simulations
(e.g. see Munshi et al. 1999, Valageas et al. 1999 and  Colombi et
al. 1996 for a detailed
description of the effect of the finite volume correction involved in their
estimation). We have used $\Phi_{\eta}(y)$ as derived above to compute
$P(\eta)$ with the help of equation (\ref{pdf}).

%Since the variance associated with $\eta$ is much smaller 
%than unity and we have to use the integral definition of $P(\eta)$ to
%recover the PDF from $\Phi_{\eta}(y)$ and the use of $h(x)$ is
%appropriate only for the limiting case of large variance.

An Edgeworth expansion of the pdf can be made by starting with
the simpler form $P(\eta({\theta_0}))$, which can then be used to 
construct the Edgeworth series for
$P(\kappa({\theta_0}))$. The Edgeworth expansion
(see e.g. Bernardeau \& Koffman 1994)
is meaningful when the variance is less than unity, which guarantees a
convergent series expansion in terms of Hermite polynomials $H_n(\nu)$,
of order n and with $\nu = \eta/\sqrt{(\xi_{\eta}}$). Hence it is
used in quasi-linear analysis with the perturbative expressions
for cumulants. However 
the $S^{\eta}_N$ parameters used in the expansion of $P(\eta)$
are from the highly 
non-linear regime; i.e. although the
variance is smaller than unity, the parameters that
characterize it are from the highly non-linear dynamics of the underlying dark
matter distribution.

\begin{equation}
P(\eta) \equiv { 1 \over  \sqrt 2 \pi {\bar \xi}_{\eta} } \exp ( - {\nu^2 \over 2} )
\Big [ 1 + {\sqrt {\bar \xi}_{\eta}} 
{S_3^{\eta} \over 6} H_3(\nu)+ {\sqrt {\bar \xi}_{\eta}}^2 \Big
( {{S_4^{\eta} \over 24} H_4(\nu)+ {{S_3^{\eta}}^2 \over 72}} H_6(\nu) \Big ) +  \dots  \Big ]
\end{equation}

\n
%For the purpose of weak lensing statistics $\Sigma_n$
%parameters are generally used instead of $S_n$ parameters. In terms of
%these parameters we can express $P(\eta)$ as:

%\begin{equation}
%P(\eta) \equiv { 1 \over  \sqrt 2 \pi {\bar \xi}_{\eta} } \exp ( - {\nu^2 \over 2} )
%\Big [ 1 + 
%{\Sigma_3^{\eta} \over 6} H_3(\nu)+  \Big
%( {{\Sigma_4^{\eta} \over 24} H_4(\nu)+ {{\Sigma_3^{\eta}}^2 \over 72}} H_6(\nu) \Big ) +  \dots  \Big ]
%\end{equation}

The magnification $\mu$ can also be used instead of $\kappa$ 
using the weak lensing relation $\mu({\theta_0}) = 1 + 2 \kappa({\theta_0})$. 
Its minimum value can be related to $\kappa_{min}$ defined earlier as
$\mu_{min} = 1 + 2 \kappa_{min}$. Finally, the reduced convergence $\eta$
and the magnification $\mu$ can be related by the following
equation (Valageas 1999):

\begin{equation}
\eta{(\theta_0)} = {\mu({\theta_0}) - \mu_{min} \over 1 - \mu_{min}} .
\end{equation} 

\n
So we can express the relations connecting the probability distribution
function for the smoothed convergence statistics
$\kappa({\theta_0})$, the reduced
convergence $\eta({\theta_0})$ and the magnification $\mu({\theta_0})$ as,

\begin{equation}
P(\kappa({\theta_0})) = 2 P(\mu) = P(\eta) { 2 \over ( 1 -  \mu_{min})
} = P(\eta) { 1 \over |\kappa_{min}|} .
\end{equation}

Throughout our analysis we have used a top-hat filter for smoothing
the convergence field, but our study can be extended to
compensated  filters (Schneider et al. 1998; Reblinsky et al. 1999),
which may be more suitable for observational purposes. 
The formalism which we have developed for one-point statistics such as
the PDF and the VPF can also be extended to compute the bias and higher order 
cumulants associated with spots in $\kappa$ maps above a certain
threshold. The statistics of such spots can be associated with the 
statistics of over-dense regions in the underlying mass distribution
which represent the collapsed objects. A detailed analysis of these
issues will be presented elsewhere (Munshi \& Coles 1999b).

\begin{figure}
\protect\centerline{
 \epsfysize = 4.truein
 \epsfbox[22 147 583 708]
 {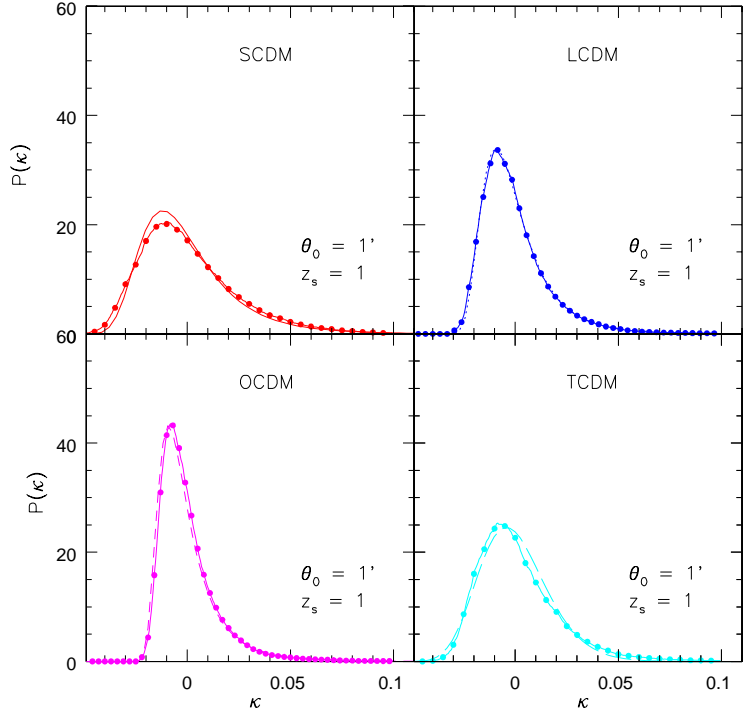} }
 \caption{Analytical predictions for probability distribution function
 of the weak lensing convergence field P($\kappa$) smoothed with an angle
 $\theta_0 = 1'$ 
 is plotted as a function of $\kappa$. The source red-shift is
 fixed at $z_s = 1$. The four curves correspond to the
SCDM, LCDM, OCDM and TCDM models as indicated. The Solid lines
 with black dots are the results from ray tracing simulations.}
\end{figure}

\begin{figure}
\protect\centerline{
 \epsfysize = 4.truein
 \epsfbox[22 147 583 714]
 {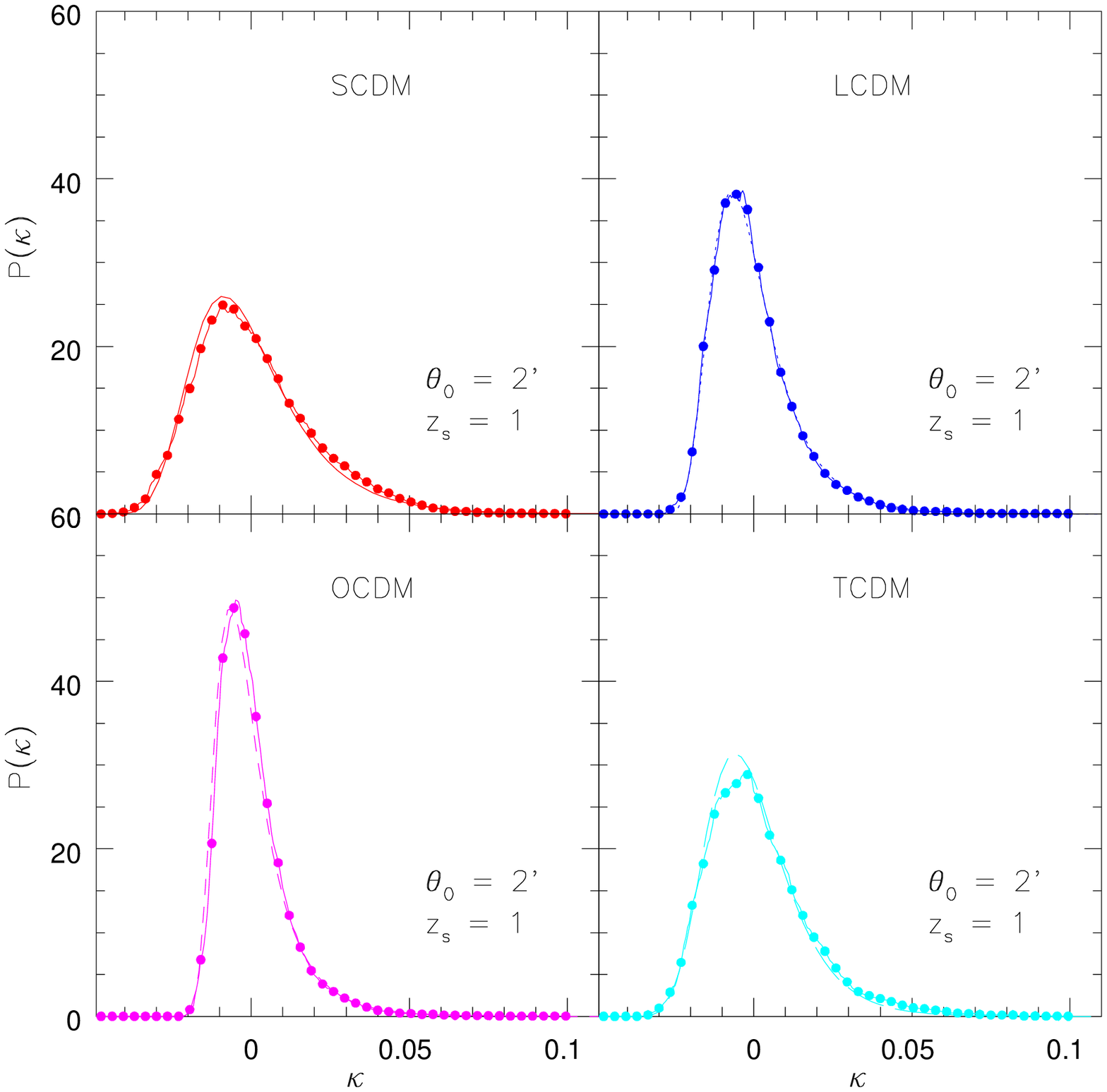} }
 \caption{As in Figure 5, with smoothing angle $\theta_0 =2'$.}
\end{figure}

\begin{figure}
\protect\centerline{
 \epsfysize = 4.truein
 \epsfbox[22 147 583 714]
 {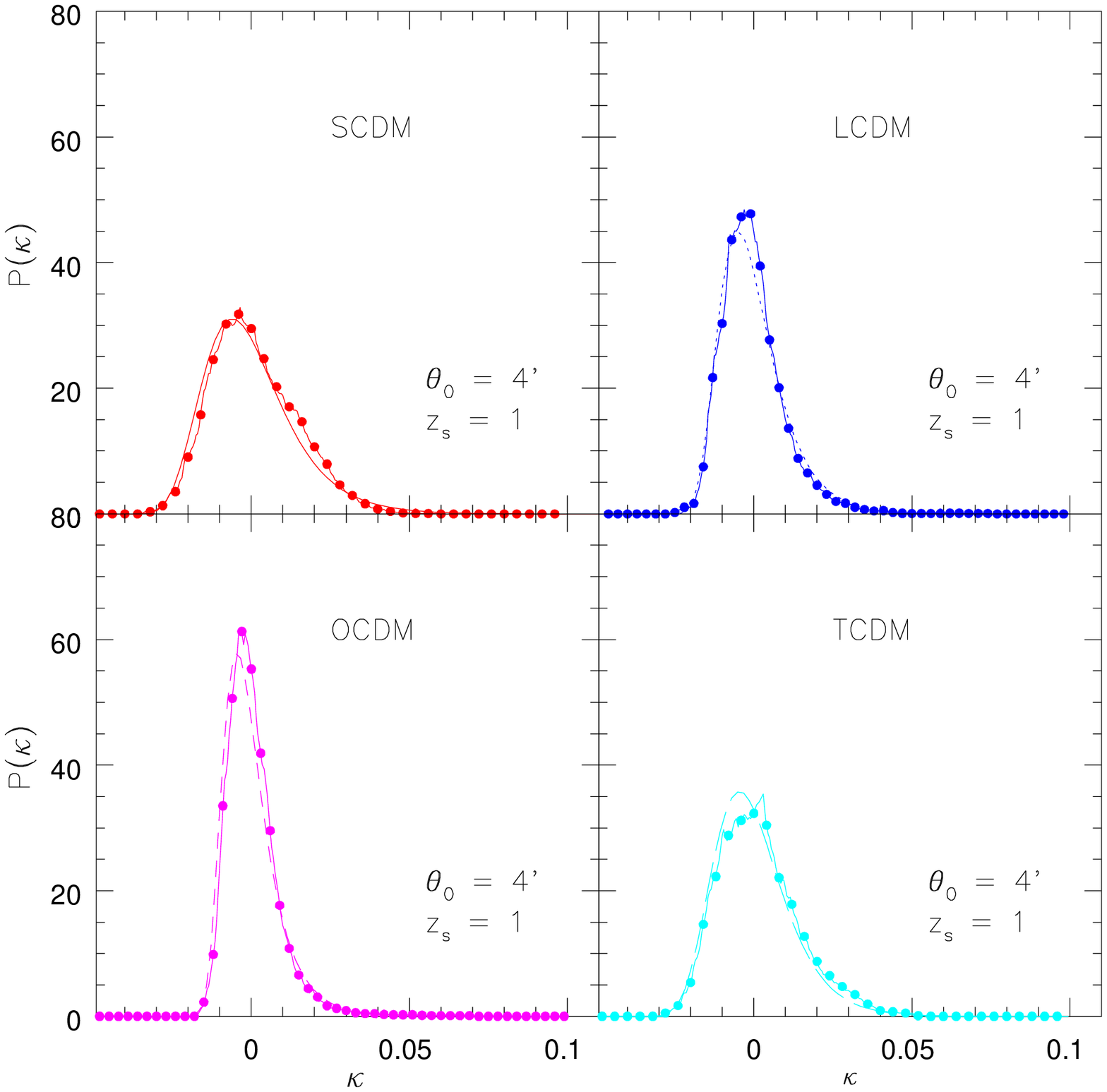} }
 \caption{As in Figure 5, with smoothing angle $\theta_0 =4'$.}
\end{figure}

\begin{figure}
\protect\centerline{
 \epsfysize = 4.truein
 \epsfbox[19 147 583 714]
 {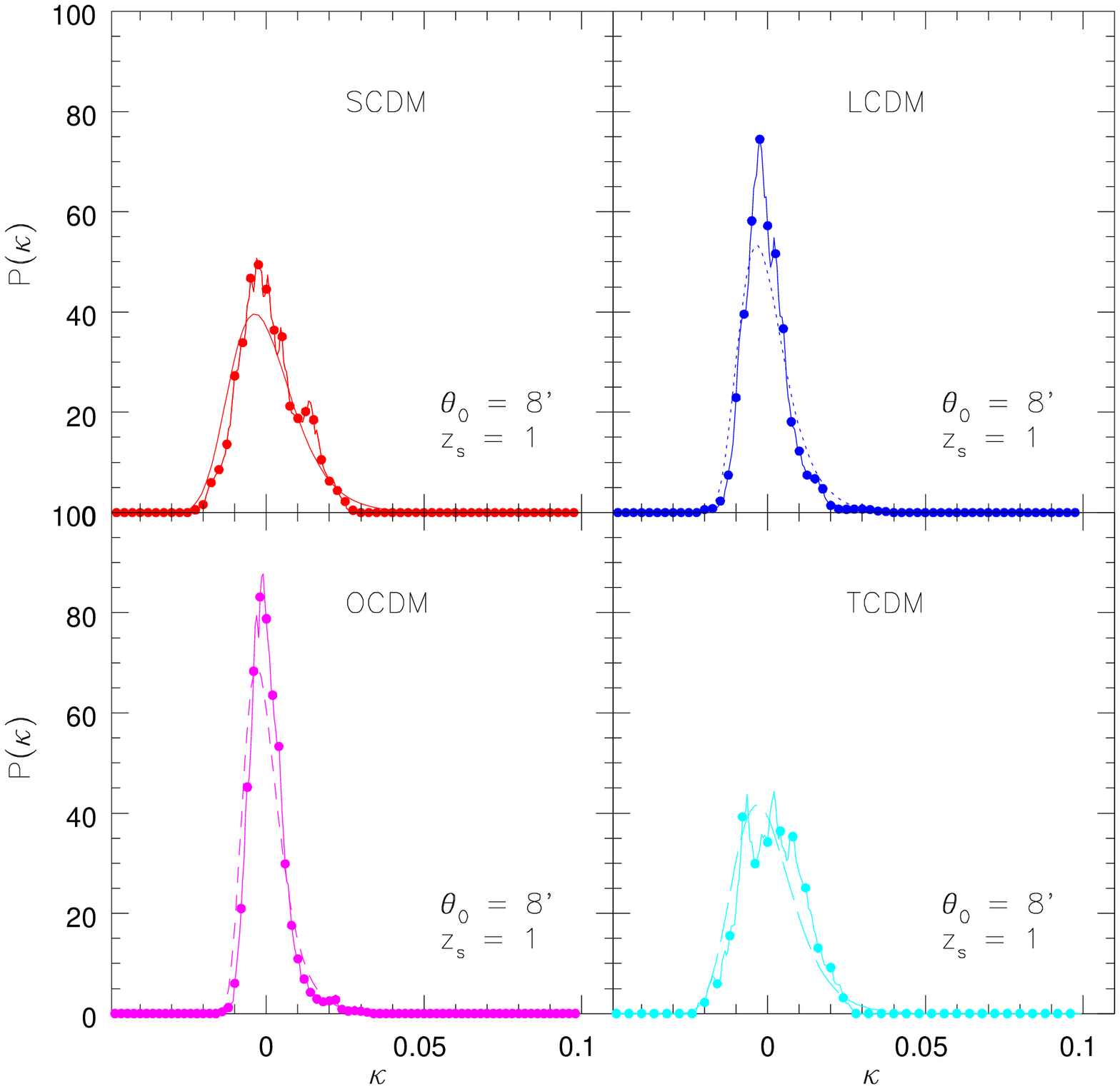} }
 \caption{As in Figure 5, with smoothing angle $\theta_0 =8'$.}
\end{figure}

\begin{figure}
\protect\centerline{
 \epsfysize = 5.truein
 \epsfbox[19 144 586 714]
 {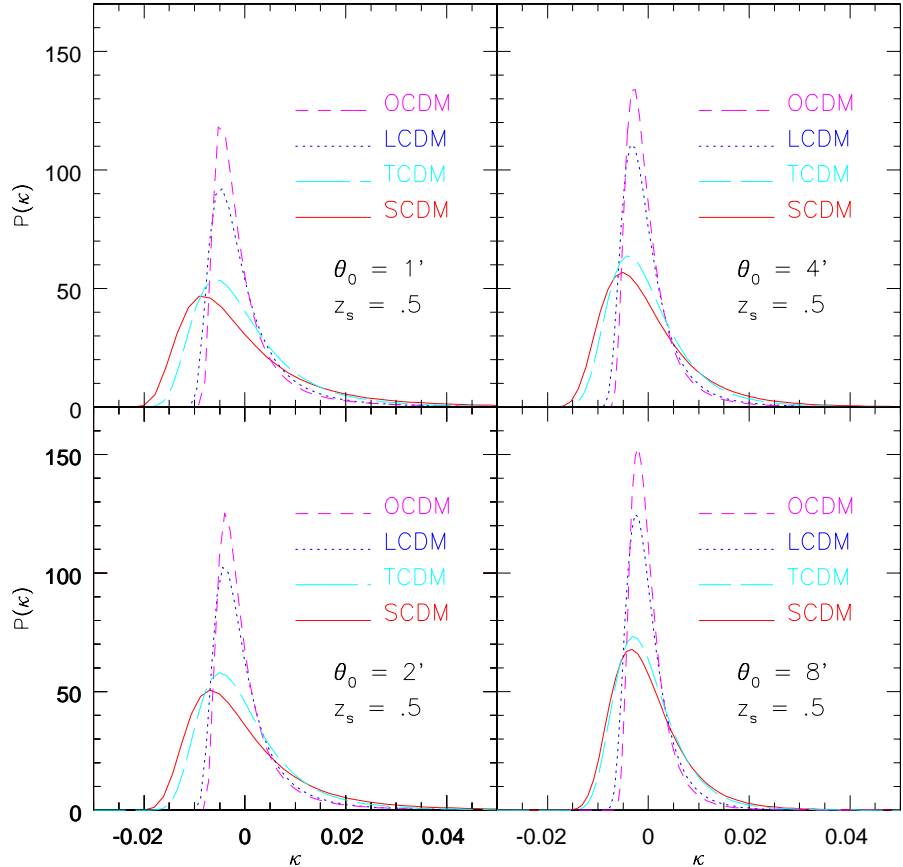} }
 \caption{Analytical predictions for the probability distribution function
 P($\kappa$), smoothed with angle
 $\theta_0$, as a function of $\kappa$. The smoothing angle
 varies from $1'$ to $8'$. The source red-shift is
 fixed at $z_s = 0.5$. The curves with increasing peak heights 
 correspond to the SCDM, LCDM, OCDM and TCDM models. }
\end{figure}

\begin{figure}
\protect\centerline{
 \epsfysize = 5.truein
 \epsfbox[19 144 586 714]
 {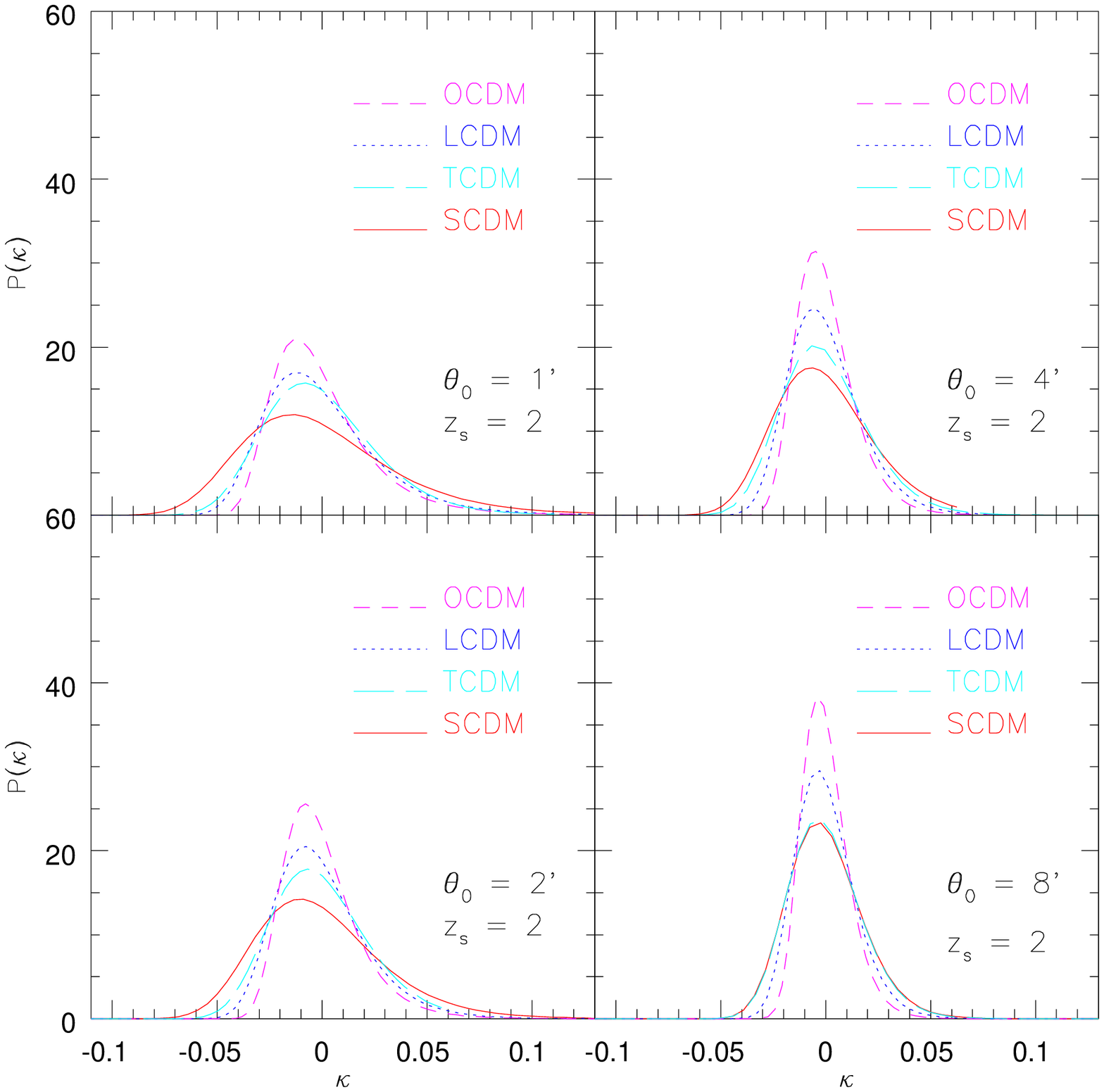} }
 \caption{As in Figure 9, with source redshift $z_s = 2$.}
\end{figure}

\section{Comparison with Ray Tracing Simulations}

To compare analytical results with numerical simulations we
smooth the convergence maps generated from numerical
simulations using a top-hat filter of suitable smoothing angle $\theta_0$.
The minimum smoothing radius we have used is $1'$ which is 
much larger than the numerical resolution length scale. 
%In our 
%earlier studies we found that numerical artifacts are not the dominant
%contribution even for angles as small as $.25$ arc-minute. 
The maximum 
smoothing radius we have studied is $8'$  which is much 
smaller than the box size and we expect that finite volume effects are
not significant in our studies. The box size is $L = 166.28'$
for the EDS models,
$L =235.68'$ for the $\Omega=0.3$ open model, and $L = 209.44'$
for the $\Omega=0.3$  model with cosmological constant $\Lambda =
0.7$. Numerical outputs from one realization of each of these
cosmological models were used to compare against theoretical results. 
Elsewhere, we have used these simulations to test analytical results
for cumulants and cumulant correlators and found very good agreement.

Figures 4-8 show the analytical and numerical pdf's for the four
cosmological models for $z_s=1$ and 
smoothing angles ranging from $\theta_0=1'-8'$. Figures
9 and 10 show the dependence on $z_s$, by showing the pdf's for 
$\theta_0=1'$ with $z_s=0.5$ and $z_s=2$ respectively. 
Several qualitative features in 
$P(\kappa(\theta_0))$ can be understood
from the results shown. 
%As we
%increase the smoothing angle, the peak of the PDF approaches
%its mean value zero and takes a  Gaussian shape.
As we decrease the smoothing angle, the peak of the pdf
curves shifts towards negative values of $\kappa(\theta_0)$, 
the peak height decreases and the distribution becomes more
non-Gaussian. The reason is that when we decrease the
smoothing angle we probe more non-linear scales in the underlying 
mass distribution and therefore the probability distribution for 
$\kappa(\theta_0)$ is more non-linear. 
Increasing the source red-shift $z_s$ has two distinct
effects. It introduces more dark matter regions
 between the observer and the source which are virtually uncorrelated
with each other.  This has the effect of making
probability distribution more Gaussian, even though the variance
increases due to the increased path length. The peak height of the PDF
decreases with increasing red-shift $z_s$ and $P(\kappa(\theta_0))$ becomes
broader with the increase in the variance.

For given smoothing angle and fixed redshift, the PDFs of the OCDM model
are peaked most sharply, followed by that of the LCDM, TCDM and SCDM
models. With the 
reduced convergence field $\eta(\theta_0)$, 
$P(\eta(\theta_0))$ does not depend on the
background geometry of the universe. 
We have noted that 
due to the presence of an extra $|\kappa_{min}|$ in the denominator, the 
variance of the reduced convergence
$\eta$ actually decreases with red-shift $z_s$ in contrast to
the variance of $\kappa(\theta_0) $. The
skewness and other higher order moments for $\eta$ are also the same as for
underlying mass distribution $\delta$. It is therefore easy to understand
that the PDF of the $\eta$ field is independent of background cosmological
parameters and is very similar to the PDF of the density contrast $\delta$. 
However the difference is that 
variance of $\delta$ is much larger than unity, while that of $\eta$
is much smaller than unity, although the moments and hence
the VPF associated with them are exactly the same. 
The cosmological dependence of $P(\kappa)$ enters through
$\kappa_{min}$ and $\langle \eta^2(\theta_0)
\rangle$. Since $P(\eta(\theta_0))/|\kappa_{min}| =
P(\kappa(\theta_0))$, the peak height of $P(\eta)$
is determined by the relative ordering of $|\kappa_{min}|$, for given
red-shift.
%(note that $|\kappa_{min}|$ is independent of smoothing angle
%$\theta_0$ and dependence of $\kappa$ PDF on smoothing angle $\theta_0$ is
%entirely dependent on variance $\langle \eta^2(\theta_0)
%\rangle$). 
Thus we find that $|\kappa_{min}|$ and $\langle
\eta^2(\theta_0)\rangle$ play a very important role in the 
construction of $P(\kappa(\theta_0))$.

The approximation we have used to simplify the void probability
distribution function of $\eta$ also gives us simple and
new powerful method to compute the $S_N$ parameters for the
convergence field, 
by relating them with with the 
$S_N$ parameters of the underlying mass distribution as
$S_N^{wl} = S_N /({|\kappa_{min}|})^{N-2}$. 
%It is possible to expand $P(\eta)$
%in an Edgeworth expansion, as is generally done for the case of
%$P(\delta)$. 

The comparison of the analytical results with numerical simulations 
shown in figures 4-8
shows that there is a very good match, 
particularly for small smoothing angles. 
The exponential tail for positive values of $\kappa$ is well reproduced
by our analytical results. The analytical PDF also reproduces
correctly the exact position of the maxima and the sharp fall of the PDF 
for negative values of $\kappa$. The relative ordering of the peaks
for various cosmological models are also well reproduced. For large 
smoothing angles there is some disagreement between the analytical
predictions and the ray tracing experiments. This is partially due to
the fact that local initial spectral index for length scales 
making the dominant contribution for larger smoothing angles is 
different compared to that for the case of small smoothing angle.
We have chosen $\omega = .3$ for all our theoretical models, 
which correspond to initial power spectral index of $n=-2$. 

For larger smoothing angles length scales contributions to the
one-point statistics of the convergence field come from 
quasi-linear length scales, hence the non-linear theory will not be
accurate. In such cases one has to change the generating function
${\cal G}(\tau)$ to the perturbative generating function ${\cal
G}^{PT}(\tau)$. With this change, we can
extend all our results in a straight forward way. From the point of view
of the Edgeworth expansion, the only change we need to
make is in the values of $S_N$ parameters, where we have to  
replace the values obtained from the hyper-extended perturbation theory with 
ones obtained from the tree-level perturbation theory.
However for very large smoothing angles ($\theta_0 >1^\circ$) we have
to keep in mind that the small angle
approximation is a necessary ingredient in all our analytic
calculations.

\section{Discussion}

We have found an analytic expression for the PDF $P(\kappa(\theta_0))$
of the smoothed convergence field $\kappa(\theta_0)$. This is a 
generalization of our earlier work on lower order
cumulants. We found very good agreement of our analytic results with
results from ray tracing experiments. We found that 
the PDF of the convergence field 
$\kappa(\theta_0)$ can
distinguish different cosmological models. We also show that
one can define a reduced convergence field $\eta(\theta_0)$ which has 
the same reduced moments as its counterpart for the density contrast $\delta$.
%The difference between these two is in their variance. Whereas 
%the density PDF corresponds to very high variance, the $\kappa$ PDF 
%corresponds to variance typically less than unity. While 
%the variance  $\langle \kappa(\theta_0)^2 \rangle$ increases with
%source redshift $z_s$, the variance $\langle \eta_{\theta_0}^2
%\rangle$ decreases with source red-shift $z_s$. 
%The PDF for $\kappa$ typically shows an exponential tail for large values 
%of $\kappa$ and a sharp cutoff for small values of $\kappa$.

While the variance of the convergence 
increases with source red-shift, the PDF itself tends to become more
Gaussian. The differences in PDFs of different cosmological models
become less prominent for very large source red-shift. From an
observational view point this means that while the detection of weak
lensing is easier from deep red-shift surveys, distinguishing cosmological 
models is easier when the source red-shift is smaller.
We have not included the effect of noise due to intrinsic source
ellipticity  in our calculations. 
Several authors have focussed on weak gravitational lensing effects
in the determination of cosmological parameters from SNeIa
observations (e.g. Valageas 1999). Our results match with their
finding for small smoothing angles.

The study relating the PDF of the convergence field 
$\kappa(\theta_0)$ with the PDF of the
density contrast $\delta$ can be easily extended to relate the
$\kappa_{\theta_0}$ field with the bias associated with high peaks of the
 density contrast field $\delta$. In recent studies 
(Munshi, Coles \& Melott 1999a,b,c; 
Bernardeau \& Scheaffer 1999) it has been shown that the generating 
function approach used here can be generalized to compute not only 
the bias associated with the over dense cells but can also provide
valuable insight into the statistics of collapsed objects such as 
cumulants and cumulant correlators. Using the same analytical tools 
as we have used here, it can be shown that the statistical properties 
of high $\kappa(\theta_0)$ spots in convergence maps, e.g. cumualnts
and cumulant
correlators associated with them, can also be related to their
counterpart for the density field $\delta$. A detailed analysis of 
the signal to noise ratio is needed to determine the feasibility of 
estimating these quantities from observational data.

%We have used a top-hat filter function for the smoothed convergence
%field $\kappa$. 
%It has been pointed out that compensated
%filters are more appropriate for observational studies of weak lensing
%distortions. It is possible to extend our results for such filters and
%we hope to present these results elsewhere.
The analytical expression for $\kappa$ used in this work is a line of sight
integration of the density field $\delta$ which relies on the 
Born approximation, i.e. it neglects higher order correction terms in the
photon propagation equation. The error introduced by such an approximation in 
the quasi-linear regime has been studied by several authors
(e.g. Schneider et al (1997). 
Perturbative calculations tend to show that this error is negligible for lower
order cumulants, clearly such an analysis is not possible in the highly
non linear regime. However since the tail of $P(\kappa(\theta_0)$
contains information on all orders, a good match between theoretical
predictions and the simulation results found by our study indicate that
such corrections are negligible even in the highly non-linear regime.

\section*{Acknowledgment}
Dipak Munshi was supported by a fellowship from the Humboldt foundation at MPA
where this work was completed. It is a pleasure for Dipak Munshi to acknowledge
many helpful discussions with Patrick Valageas, Peter Coles and Katrin
Reblinsky. The complex integration routine used to generate 
$P(\eta)$ was made available to us by Francis Bernardeau; we are
grateful to him for his help.

\end{document}